\documentclass[%
 reprint,
 superscriptaddress,
 amsmath,amssymb,
 aps,
 prb,
]{revtex4-1}

\usepackage{graphicx}
\usepackage{dcolumn}
\usepackage{bm}
\usepackage{hyperref}
\usepackage[hang]{subfigure}
\usepackage[toc,page]{appendix}

\begin{document}

\title{Magnetic Skyrmion Lattice by Fourier Transform Method}

\author{Eugene Balkind}
\affiliation{Royal Holloway, University of London, TW20 0EX Egham, United Kingdom}
\author{Aldo Isidori}
\affiliation{Royal Holloway, University of London, TW20 0EX Egham, United Kingdom}
\affiliation{International School for Advanced Studies (SISSA), Via Bonomea 265, I-34136 Trieste, Italy}
\author{Matthias Eschrig}
\affiliation{Royal Holloway, University of London, TW20 0EX Egham, United Kingdom}

\date{\today}

\begin{abstract}
We demonstrate a fast numerical method of theoretical studies of skyrmion lattice or spiral order in magnetic materials with Dzyaloshinsky-Moriya interaction. The method is based on the Fourier expansion of the magnetization combined with a minimization of the free energy functional of the magnetic material in Fourier space, yielding the optimal configuration of the system for any given set of parameters. We employ a Lagrange multiplier technique in order to satisfy micromagnetic constraints. We apply this method to a system that exhibits, depending on the parameter choice, ferromagnetic, skyrmion lattice, or spiral (helical) order. Known critical fields corresponding to the helical-skyrmion as well as the skyrmion-ferromagnet phase transitions are reproduced with high precision. 
Using this numerical method we predict new types of excited (metastable) states of the skyrmion lattice, which may be stabilized by coupling the skyrmion lattice with a superconducting vortex lattice.  The method can be readily adapted to other micromagnetic systems.
\end{abstract}

\maketitle

\section{\label{sec:level1} INTRODUCTION}

Apart from the more familiar ordered magnetic phases, such as ferro- or antiferromagnetism, in recent years a plethora of more complex magnetically ordered states has been unveiled, both theoretically and experimentally. Among these states, a crucial role is played by magnetic skyrmions,\cite{Skyrme61} characterized by a whirling pattern of localized Heisenberg spins. \cite{Bogdanov89,Bogdanov89_SovPhys,Bogdanov94,Bogdanov94b,Barrett95,Brey95,Bogdanov99,Gervais05,Roessler06,Muhlbauer09,Neubauer09,Munzer10,Yu10,Nagaosa13,Romming13,Romming15,Leonov16,Leonov17} Magnetic skyrmions can be regarded as localized topological defects of the ferromagnetic state that can only be moved around or deformed in shape and size, but cannot be removed by a smooth deformation of the magnetization pattern due to micromagnetic constraints.
They either exist as isolated defects, as, for example, in certain types of topological domain walls found in helical magnets, or they themselves give rise to an extended, energetically favorable phase of helical magnets, in the form of a skyrmion lattice.

Skyrmion lattices were first introduced 
by Klebanov \cite{Klebanov85} for neutron crystals, whereas Bogdanov and Yablonskii \cite{Bogdanov89} discussed the possibility of thermodynamically stable magnetic vortices (now called skyrmions lattices) arising in anisotropic ferromagnets as an intermediate phase between the uniform and the spiral magnetic order.
Specific theoretical predictions for skyrmion lattices in, e.g., $\rm{MnSi}$, $\rm{FeGe}$, and $\rm{Fe_{1-x}Co_xSi}$,
where made by Bogdanov and Hubert,\cite{Bogdanov94} and subsequently
skyrmion lattices were reported experimentally 
in $\rm{MnSi}$ \cite{Muhlbauer09, Neubauer09} 
and $\rm{Fe_{1-x}Co_xSi}$.\cite{Munzer10, Yu10} 
The earliest theoretical predictions \cite{Bogdanov94} and experimental evidences \cite{Nagaosa13} showed that magnetic skyrmions typically form 2D triangular lattices that are perpendicular to an external magnetic field and translation invariant in the parallel direction. More recent theoretical \cite{Lin15,Rowland16} and experimental \cite{Kurumaji17,Yu18} studies have eventually revealed that an easy-plane magnetic anisotropy may also lead to the stabilization of square skyrmion and meron-antimeron \cite{Gross78,Affleck86,Moon95,Brey96} (half-skyrmion) lattices.

The formation of magnetic skyrmions may be attributed to various microscopic mechanisms, which often cooperate with each other. We can nevertheless classify these mechanisms into two broad classes, according to the relative size of the skyrmions with respect to the microscopic lattice constant. In the first class of mechanisms, skyrmions arise from the competition between the ferromagnetic exchange coupling and an anisotropic exchange interaction originating from relativistic spin-orbit coupling, namely the so-called Dzyaloshinsky-Moriya (DM) interaction,\cite{Dzyaloshinsky58,Dzyaloshinskii64,Moriya60} which breaks either inversion or mirror symmetry. 
While the ferromagnetic exchange tends to align all spins in the same direction, DM interaction favors a non-collinear alignment, thereby twisting neighboring spins. The size of a spiral or skyrmion pattern emerging from this mechanism is typically two orders of magnitude larger than the crystal lattice constant. Hence, in this case skyrmions are topologically robust against lattice defects and can be very well described in the continuum limit approximation. In the second class of mechanisms, instead, a skyrmion pattern may arise as a result of competing ferro- and antiferromagnetic exchange interactions in frustrated magnets,\cite{Okubo12,Leonov15,Lin16,Hu17} or in the presence of four-spin interactions.\cite{Heinze11} The main distinction from the first class of skyrmion patterns is the length scale of the spin modulation, which in this case is of the order of the underlying lattice constant, leading to atomic-scale skyrmion patterns that are typically commensurate with the crystal lattice. Note that this type of skyrmions do not require a broken inversion or mirror symmetry as in the case of DM chiral magnets.

In this work we focus on the large-scale, topologically robust, skyrmion patterns that characterize chiral magnets. Moreover, we consider strictly 2D materials as realized, e.g., in thin films of $\rm{Fe_{1-x}Co_xSi}$ or $\rm{FeGe}$.\cite{Yu10,Huang12} Indeed, in 3D chiral magnets the skyrmion lattice phase is thermodynamically stable only at finite temperatures and can be stabilized down to zero temperature only by a strong easy-axis anisotropy\cite{Butenko10} or the presence of multiple types of spin-orbit couplings.\cite{Rowland16} Instead, in 2D materials the skyrmion phase is stable at zero temperature over a wide range of external magnetic fields even in the absence of magnetic anisotropies.

The model Hamiltonian for Heisenberg spins in the presence of a DM interaction and an external magnetic field is given by, 
	\begin{equation}
	\hat{H}_M = - J \sum_{\left\langle ij \right\rangle} \vec{S}_i \cdot \vec{S}_j - \sum_{ij} \vec{D}_{ij} \cdot \left(\vec{S}_i \times \vec{S}_j \right) + g \mu_B \sum_j \vec{S}_j \cdot \vec{B} , \label{eq:SpinHamiltonian}
	\end{equation}
	where $J$ is the ferromagnetic exchange constant, $\vec{D}_{ij}$ is the DM coupling vector, $\vec{B}$ is the external field, $\mu_B$ is Bohr's magneton, and $\vec{S}_i$ is a Heisenberg spin at site $i$. 
The first term in Eq.~(\ref{eq:SpinHamiltonian}) represents the Heisenberg exchange interaction, the second term the anisotropic DM interaction, and the last term the Zeeman energy.
The specific form of the DM vector $\vec{D}_{ij}$ depends on the type of relativistic spin-orbit coupling present in the system.\cite{Rowland16} A Dresselhaus spin-orbit coupling results in a broken inversion symmetry, $\vec{r} \to - \vec{r}$, and a DM vector that is parallel to the relative direction $\hat{r}_{ij} \equiv (\vec{r}_{i} - \vec{r}_{j})/|\vec{r}_{i} - \vec{r}_{j}|$ of neighboring spins, $\vec{D}_{ij} = D_{\parallel} \hat{r}_{ij}$. This induces a rotation of the spin direction within the plane that is perpendicular to the inter-spin direction $\hat{r}_{ij}$, leading to a Bloch-like spiral or skyrmion pattern. 
Instead, a Rashba spin-orbit coupling results in a broken mirror symmetry with respect to a given plane, e.g., $z \to -z$. In this case the DM vector is perpendicular to both the broken symmetry axis and the inter-spin direction, $\vec{D}_{ij} = D_{\perp} \hat{z} \times \hat{r}_{ij}$, and the spins rotate within the plane that is simultaneously parallel to the inter-spin direction and the broken symmetry axis, leading to N\'eel-like spirals or skyrmions.
In this paper we consider only the Dresselhaus DM interaction. However, we would like to stress that the method presented here is completely general in this respect and can be readily modified, at no additional numerical cost, to include both types of DM interactions and any easy-axis or easy-plane magnetic anisotropy.


	%
	\begin{figure}[!tb]
	\centering
	{\includegraphics[width=0.45\textwidth]{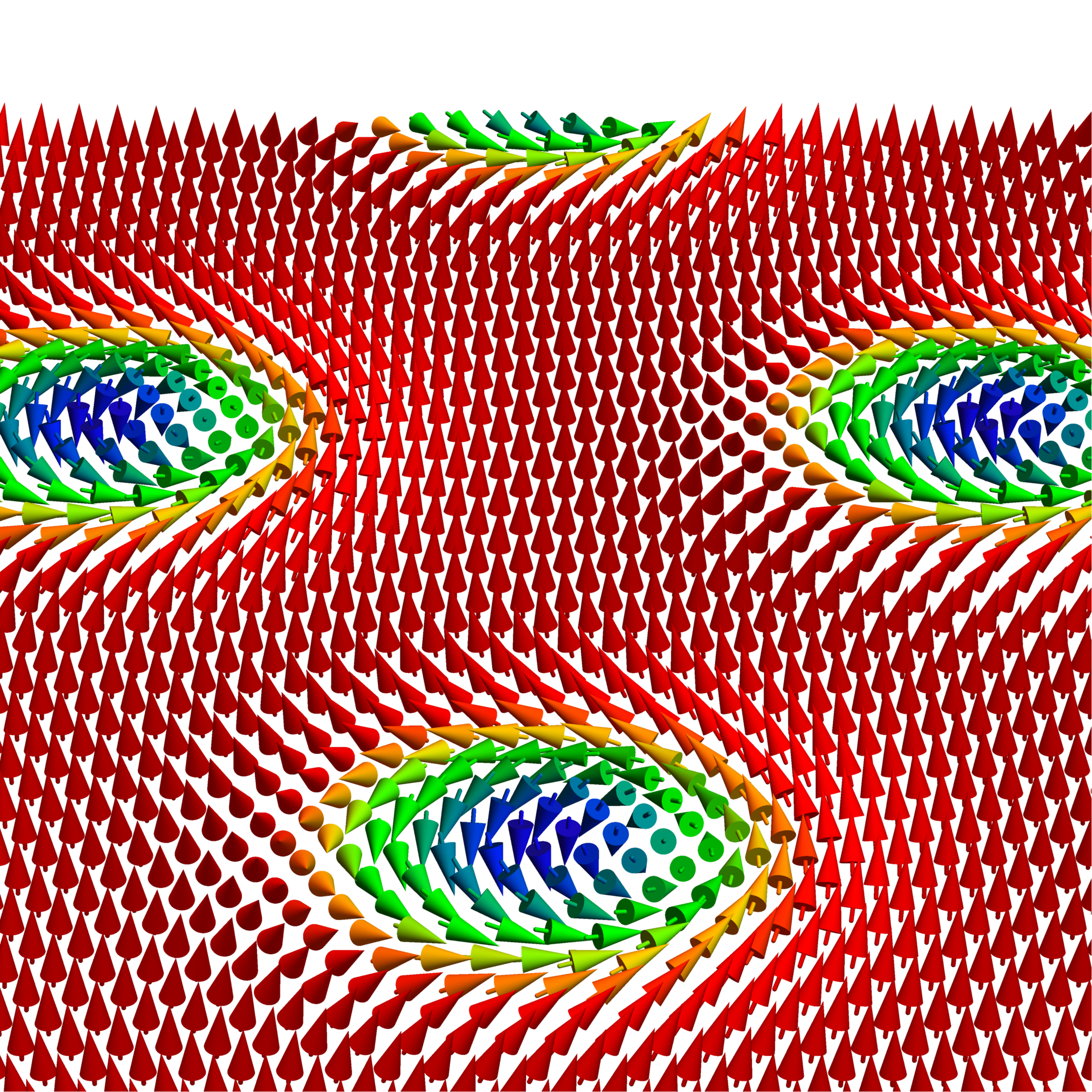}}
	\caption{Skyrmions spin texture in a triangular lattice: Spin changes from ``down'' position in the center of a skyrmion to ``up'' position far away from the center of a skyrmion.}
	\label{Intro}
	\end{figure}

	In the absence of an external magnetic field (or if the field is sufficiently weak so that the Zeeman energy can be neglected), the interplay of DM and exchange interactions leads to a helical (spiral) spin order.\cite{Bak80} 
For higher fields the Zeeman interaction becomes relevant and the spiral order gives way to a more complex magnetic state (the skyrmion lattice) until eventually at highest fields a ferromagnetic state is stabilized.
Skyrmion magnetic order (see Fig.~\ref{Intro}) occurs at intermediate external fields as
a compromise between DM interaction and Zeeman energy. 
In the center of a skyrmion, the spin points in the direction opposite to the magnetic field. 
When moving radially from the center towards the periphery, the direction of the spin rotates in a similar fashion as in the helical state, until it points in the direction of the magnetic field at the skyrmion boundary. 
In the regions between the skyrmions the spins stay aligned with the external magnetic field.
Hence, by varying the inter-skyrmion distance it is possible to arrive at a net energy gain resulting from
both the DM and Zeeman interactions, stabilizing a skyrmion lattice as opposed to a spiral or ferromagnetic order.

The goal of this paper is to develop a fast and accurate numerical procedure for the calculation of skyrmion lattice or helical order by utilizing a method similar to the one used by Brandt for the effective numerical implementation of superconducting vortex lattices.\cite{Brandt95}
We additionally increase the speed of the numerical convergence by invoking a virial theorem in order to find the equilibrium lattice constants for the skyrmion lattice.
This general procedure also allows for the study of metastable solutions and can be adapted to combine skyrmion lattices with superconducting vortex lattices. Such a combined system may stabilize new types of skyrmion lattices that otherwise would be metastable. 

In Section~\ref{sec:OptSkyrm} we present our analytical and numerical approach to the skyrmion lattice problem, including the results obtained for the phase diagram of a 2D non-centrosymmetric ferromagnet and the field dependence of the inter-skyrmion distance; in Section~\ref{sec:Meta} we discuss alternative metastable solutions for the skyrmion lattice and the possible ways to stabilize them, focusing in particular on the honeycomb skyrmion lattice. 
	 
	\section{\label{sec:OptSkyrm} SKYRMION LATTICE PHASE}

	\subsection{\label{sec:level2} Euler-Lagrange Equations for a Magnetic System in Fourier Space}

	In this section we write the free energy functional for a two-dimensional non-centrosymmetric magnetic system in an external field and derive the corresponding Euler-Lagrange equations for the magnetization. Afterwards, we Fourier transform the obtained equations and solve them in Fourier space.

	We start with the free energy functional written in the continuum limit. This approximation is legitimate as long as the magnetization varies on a length scale much larger than the crystal lattice spacing $a_0$, i.e., when $ \left| \nabla \vec{M} \right| a_{0} \ll 1$. In Cartesian coordinates, the free energy per unit area reads
	\begin{eqnarray}
	F\left[ \vec{M} \right] & = & \int \left\lbrace \frac{J}{2} \sum_\mu \left(\partial_\mu \vec{M} \right) \cdot \left(\partial_\mu \vec{M} \right) \right. \nonumber \\
	 & + & \left. D \vec{M}\cdot\left( \nabla \times \vec{M}\right) - \vec{B} \cdot \vec{M} \phantom{\sum_\mu \!\!\!\!\!\!\!\!} \right\rbrace \frac{dx dy}{A}, \label{Functional_Begin}
	\end{eqnarray}
	where $J$ is the ferromagnetic exchange constant, $D$ is the DM interaction, and $\vec{B}$ is the external magnetic field applied (note that, in the continuum limit, the couplings $J$, $D$, and $\vec{B}$ have different dimensions than their corresponding lattice counterparts figuring in Eq.~(\ref{eq:SpinHamiltonian}); moreover, they incorporate, as a multiplicative factor, the microscopic lattice coordination number). Here, $\vec{M} = \vec{M}(\vec{r}) = \left( M_x(x,y) , M_y(x,y), M_z(x,y) \right)$, obeying $| \vec{M} |^2 = 1$, and $\vec{B} = \left(0,0,B\right)$, with a constant $B$ corresponding to a uniform external field in the $z$-direction. We use the notation $\kappa = \frac{D}{2J}$ for the ratio between the DM interaction and the ferromagnetic exchange coupling. As $\kappa$ has dimensions of an inverse length, dimensionless coordinates result from 
$\left(\tilde{x}, \tilde{y} \right) = \left(\kappa x, \kappa y \right) = \kappa \vec{r}$. This leads to the dimensionless form of the free energy functional $\tilde{F} = F/(2 J \kappa^2)$,\cite{Han10}
\begin{eqnarray}
\tilde{F}\left[ \vec{M} \right] & = & \int \left\lbrace \frac{1}{4} \sum_\mu \left(\tilde{\partial}_\mu \vec{M} \right) \cdot \left(\tilde{\partial}_\mu \vec{M} \right) \right. \label{Functional_dimless} \\
& & + \left. \vec{M} \cdot \left(\tilde{\nabla} \times \vec{M}\right) - \vec{\beta}\cdot \vec{M} \phantom{\sum_\mu \!\!\!\!\!\!\!\!} \right\rbrace \frac{d\tilde{x} d\tilde{y}}{\tilde{A}} \nonumber \\
& = & \int \tilde{\mathcal{F}} \left[\vec{M} (\tilde{x}, \tilde{y}) \right] \frac{d\tilde{x} d\tilde{y}}{\tilde{A}} , \nonumber  
\end{eqnarray}
where $\vec{\beta} = \vec{B}/(2 J \kappa^2)$ is the dimensionless magnetic field. For simplicity, we shall drop all tildes from now on, remembering that lengths are measured in units of $\kappa^{-1}$ and energies in units of $2 J$.

Since the magnetization obeys the micromagnetic constraint $| \vec{M}(x,y) |^2 = 1$ at every point in space, we introduce a Lagrange multiplier field, $\lambda(x,y)$, in order to enforce the aforementioned condition. The total functional density then reads
\begin{eqnarray}
{\mathcal{F}} & = &\frac{1}{4} \sum_\mu \left(\partial_\mu \vec{M} \right) \cdot \left(\partial_\mu \vec{M} \right) + \vec{M} \cdot \left(\nabla \times \vec{M}\right) - \vec{\beta}\cdot \vec{M} \nonumber \\
& + & \lambda \left(|\vec{M}|^2 - 1 \right) . \label{Functional_Lagrange}
\end{eqnarray}

We now write the Euler-Lagrange equations for each component of the magnetization, $M_\mu$,
\begin{equation}
\frac{\partial \mathcal{F}}{\partial M_\mu} - \partial_\nu \frac{\partial \mathcal{F}}{\partial \left(\partial_\nu M_\mu \right)} = 0 ,
\end{equation}
with summation convention applied for $\nu$, with $\nu \in \left\{ x, y\right\}$. Explicitly, this leads to
\begin{equation}
\left(\partial_x ^2 + \partial_y ^2 \right) M_x - 4 \partial_y M_z - 4 \lambda M_x = 0 , \label{EL_x}
\end{equation}
\begin{equation}
\left(\partial_x ^2 + \partial_y ^2 \right) M_y + 4 \partial_x M_z - 4 \lambda M_y = 0 , \label{EL_y}
\end{equation}
\begin{equation}
\left(\partial_x ^2 + \partial_y ^2 \right) M_z - 4  \left(\partial_x M_y - \partial_y M_x \right) - 4 \lambda M_z = -2 \beta . \label{EL_z}
\end{equation}

Being inspired by Brandt's approach for Abrikosov vortices in type-II superconductors, \cite{Brandt95} we Fourier transform the magnetization components (and the Lagrange multiplier) and solve the problem in Fourier space. As the types of magnetic order promoted by the DM interaction, namely the skyrmion lattice and helical order, are periodic structures, the Fourier approach is preferable.
The discrete Fourier transform of $M_x$ is written as
\begin{equation}
M_x (x,y) = \sum_{m,n} X_{mn}  e^{-i\vec{k}_{mn}\vec{r}} ,
\end{equation}
where $X_{mn} = X_{\vec{k}_{mn}}$ is the Fourier coefficient for $M_x$, and the discrete wave vectors are given by 
\begin{equation}
\vec{k}_{mn} = \frac{2\pi}{x_1 y_2} \left(\begin{array}{c}
m y_2 \\
nx_1 - mx_2
\end{array}\right) = \left(\begin{array}{c}
k_x \\
k_y
\end{array} \right) , \label{eq:k-vectors}
\end{equation}
with $m$ and $n$ integer indices and, for a triangular lattice with lattice spacing $a$, $x_1 = a$, $x_2 = \frac{a}{2}$, $y_2 = \frac{a \sqrt{3}}{2}$. In order to shorten the notation, we write
\begin{equation}
M_x = \sum_{\vec{k}} X_{\vec{k}} e^{-i\vec{k}\vec{r}}, \label{Fourier_Mx}
\end{equation}
and by analogy for the remaining components of the magnetization. For the Lagrange multiplier
we write
\begin{equation}
\lambda = \sum_{\vec{k}} \lambda_{\vec{k}} e^{-i\vec{k}\vec{r}}. \label{Fourier_lambda}
\end{equation}
The Fourier transformed Euler-Lagrange equations, (\ref{EL_x})--(\ref{EL_z}), then read
\begin{equation}
\sum_{\vec{k}^\prime} \left[ \left\lbrace- k^{\prime 2} X_{\vec{k}^\prime} + 4 i k_y ^\prime Z_{\vec{k}^\prime} \right\rbrace \delta_{\vec{k}\vec{k}^\prime} - 4 X_{\vec{k}^\prime} \lambda_{\vec{k}-\vec{k}^\prime} \right] = 0 , \label{EL_Fourier_1}
\end{equation}
\begin{equation}
\sum_{\vec{k}^\prime} \left[ \left\lbrace- k^{\prime 2} Y_{\vec{k}^\prime} - 4 i k_x ^\prime Z_{\vec{k}^\prime} \right\rbrace \delta_{\vec{k}\vec{k}^\prime} - 4 Y_{\vec{k}^\prime} \lambda_{\vec{k}-\vec{k}^\prime} \right] = 0 ,\label{EL_Fourier_2}
\end{equation}
\begin{multline}
\sum_{\vec{k}^\prime} \left[ \left\lbrace- k^{\prime 2} Z_{\vec{k}^\prime} + 4 i \left( k_x ^\prime Y_{\vec{k}^\prime} - k_y ^\prime X_{\vec{k}^\prime} \right) \right\rbrace \delta_{\vec{k}\vec{k}^\prime} - 4 Z_{\vec{k}^\prime} \lambda_{\vec{k}-\vec{k}^\prime} \right] \\ = -2 \beta \delta_{\vec{k},0} . \label{EL_Fourier_3}
\end{multline}

	\begin{figure*}[!tb]
	\centering
	{\includegraphics[width=0.45\textwidth]{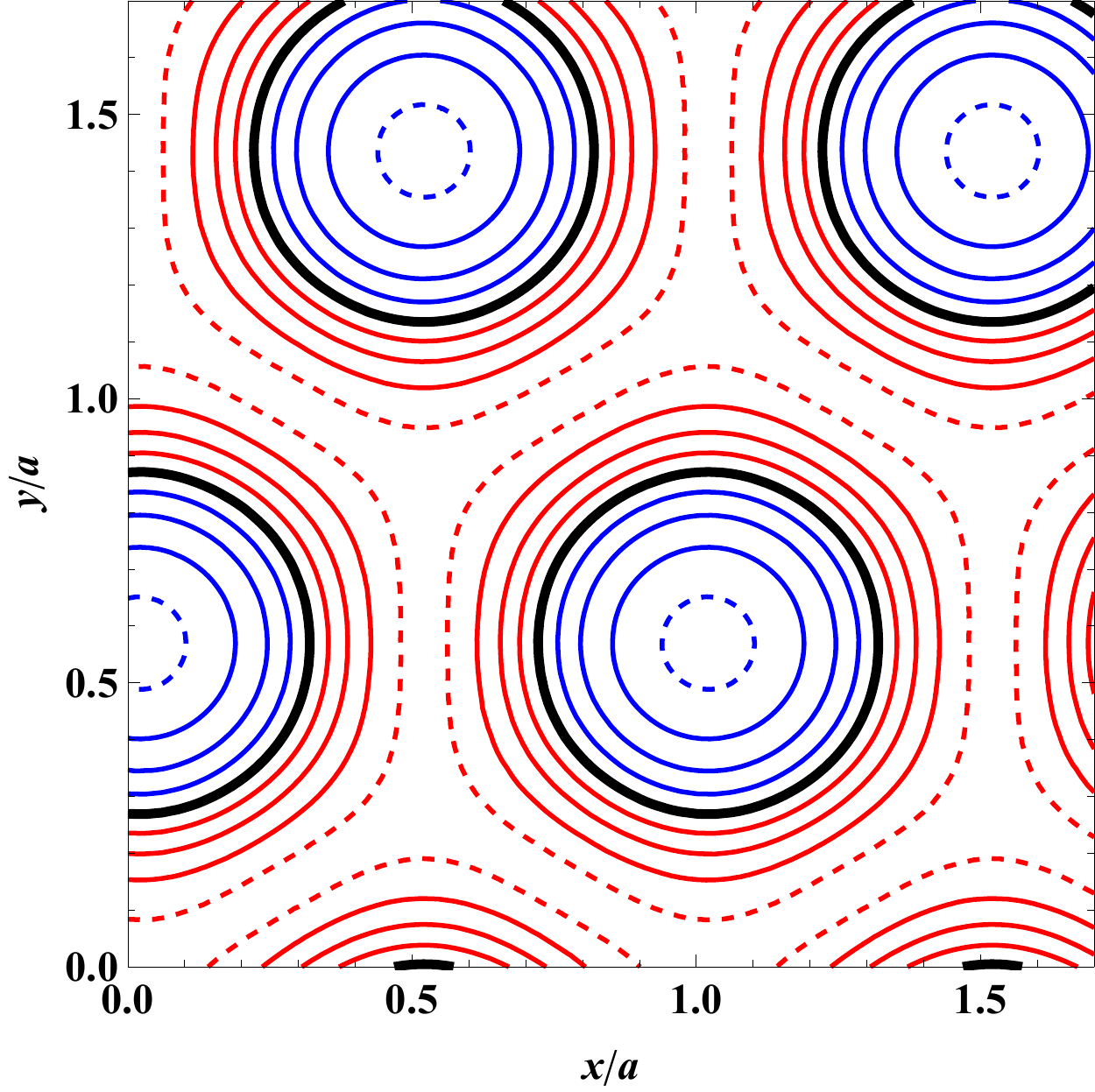}} \hfill
	{\includegraphics[width=0.45\textwidth]{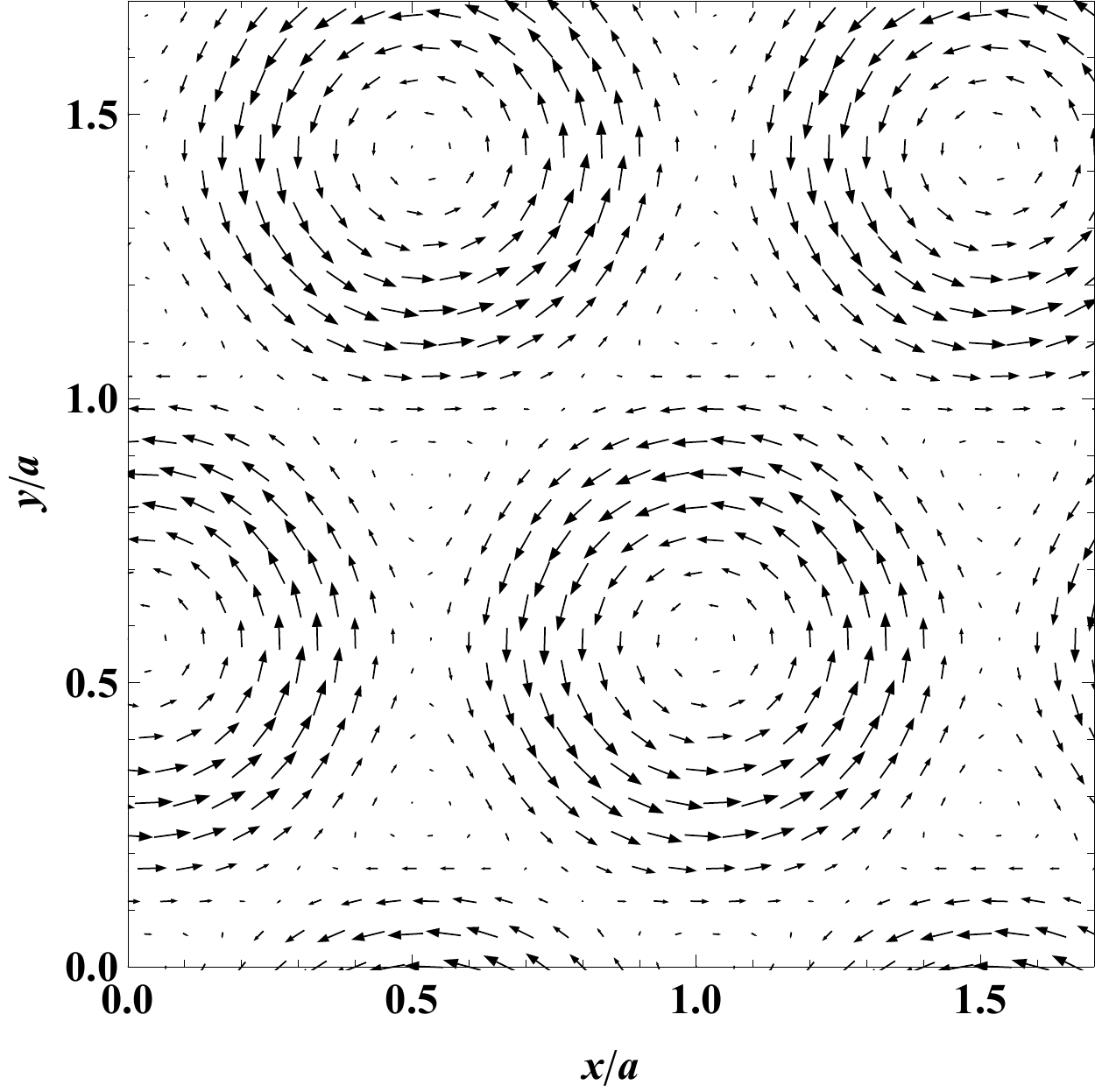}}\\[0.5cm]
	{\includegraphics[width=0.45\textwidth]{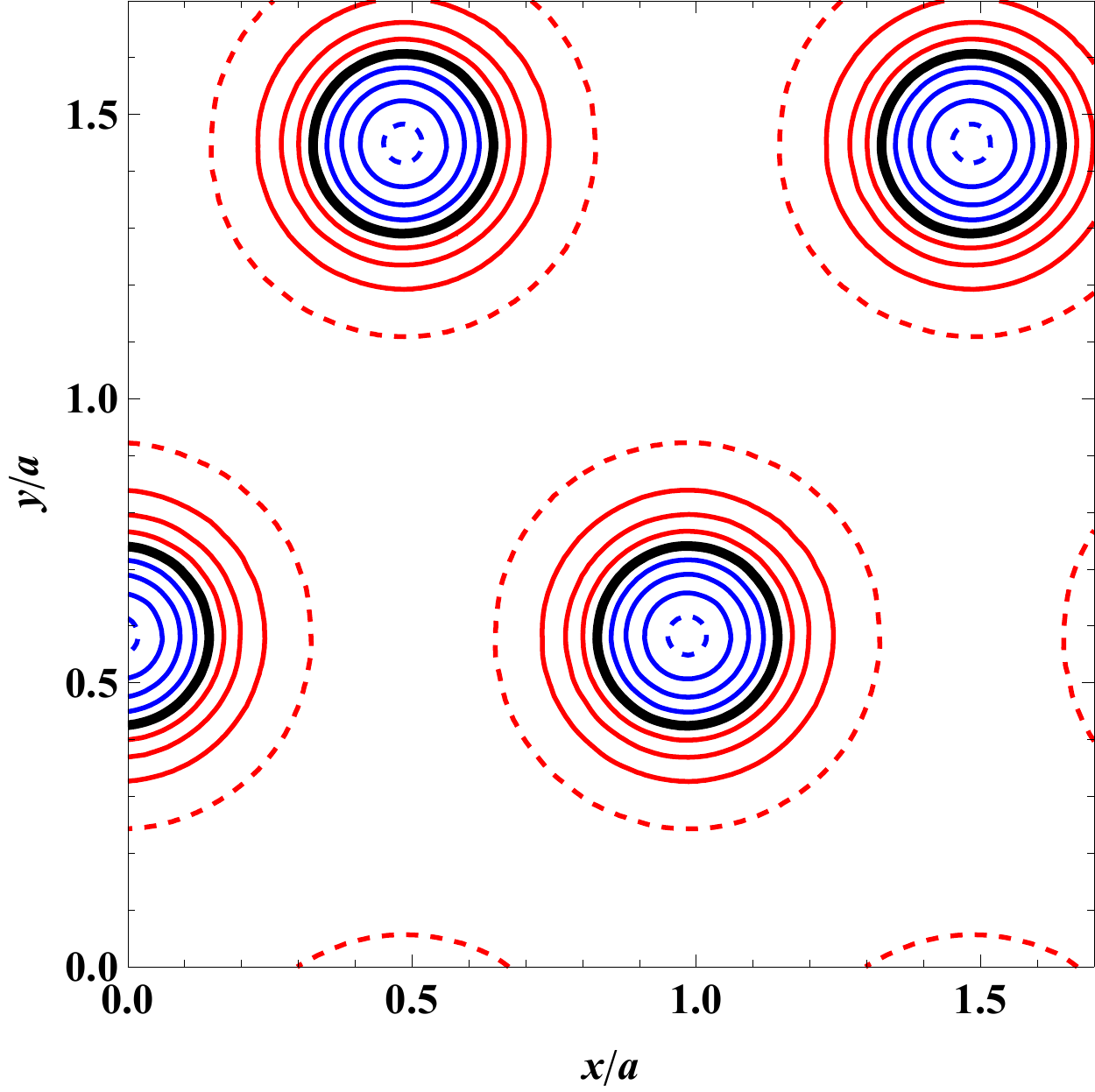}} \hfill
	{\includegraphics[width=0.45\textwidth]{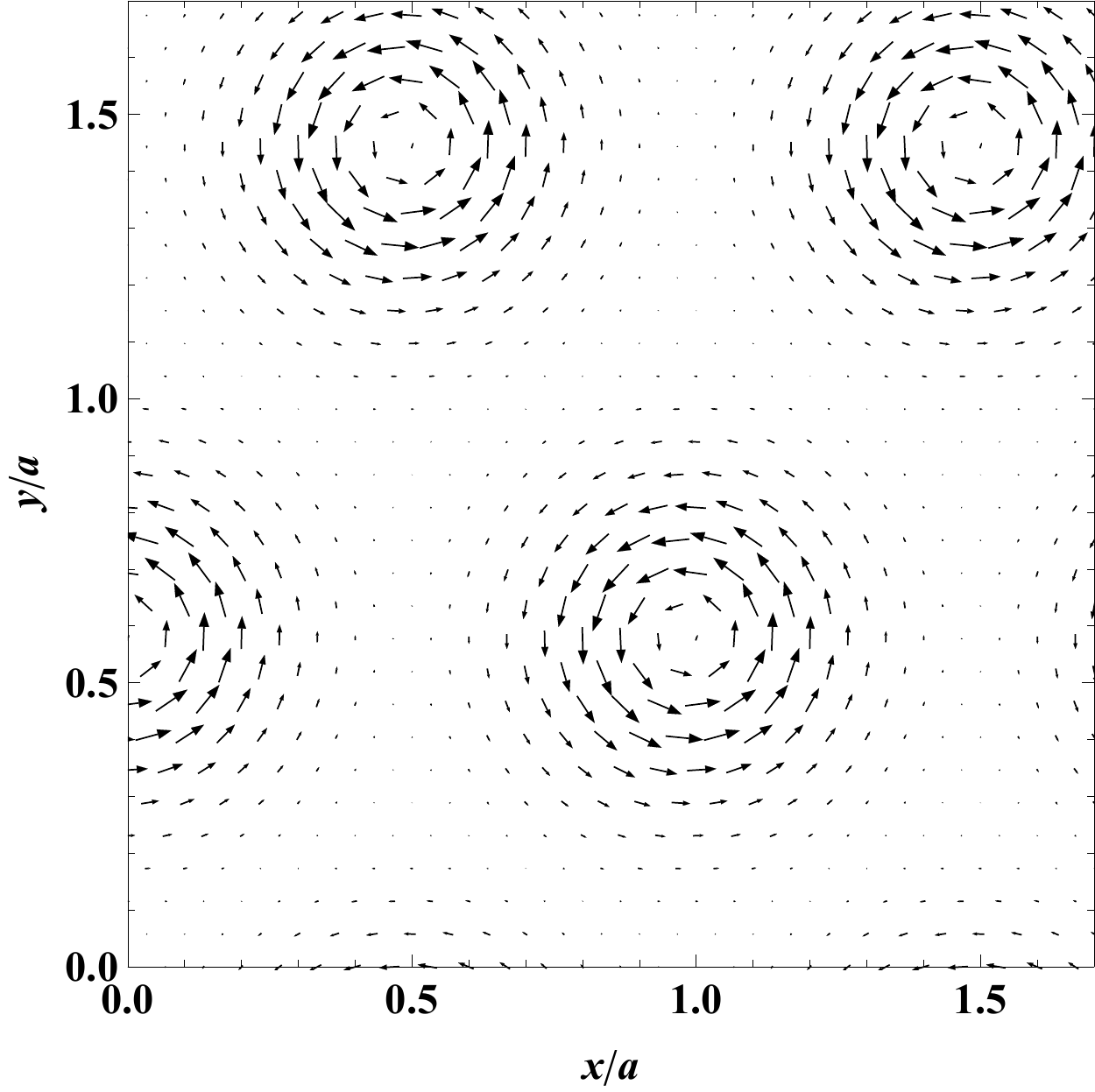}}
	\caption{
Triangular skyrmion lattice solutions. Left: contour plots for $M_z$, with the contour lines at values $-0.95$ (blue dashed), $-0.75, -0.5, -0.25,$ 0 (thick line), 0.25, 0.5, 0.75, 0.95 (red dashed), with negative values in blue and positive values in red. The centers of the skyrmions have $M_z=-1$. Right: the corresponding vector fields $(M_x,M_y)$, showing the anti-clockwise curling of the magnetization vector for each skyrmion in the lattice.
Top panels are for an external field of $\beta = 0.478$ and a lattice spacing of $a = 3.785$, corresponding to a lattice solution close to the phase transition to spiral order; 
bottom panels are for an external field of $\beta = 1.55$ and a lattice spacing of $a = 4.801$, corresponding to a lattice solution close to the phase transition to ferromagnetic order.
All spatial coordinates are normalized to the (self-consistently determined) lattice spacing $a$.}
	\label{Stable}
	\label{Mathematica_UC}
	\end{figure*}

The above equations can be rewritten in matrix notation as
\begin{equation}
\sum_{\vec{k}^\prime} \hat{D}_{\vec{k}\vec{k}^\prime} \vec{M}_{\vec{k}^\prime} = \vec{I}_{\vec{k}} , \label{Matrix_Eqn}
\end{equation}
where the term on the right hand side is proportional to the external magnetic field,
\begin{equation}
\vec{I}_{\vec{k}} = \left(\begin{array}{c}
0 \\
0 \\
-2\beta \delta_{\vec{k},0}
\end{array} \right),
\end{equation}
$\vec{M}_{\vec{k}}$ is the vector containing the Fourier coefficients of the magnetization components for a given $k$,
\begin{equation}
\vec{M}_{\vec{k}} = \left(\begin{array}{c}
X_{\vec{k}} \\
Y_{\vec{k}} \\
Z_{\vec{k}}
\end{array} \right) ,
\end{equation}
and the matrix $\hat{D}_{\vec{k}\vec{k}^\prime}$ is given by
\begin{equation}
\hat{D}_{\vec{k}\vec{k}^\prime} = \hat{K}_{\vec{k}^\prime} \delta_{\vec{k}\vec{k}^\prime} - 
4 \lambda_{\vec{k}-\vec{k}^\prime} \hat{I},
\end{equation}
with $\hat{I}$ being an identity $3 \times 3$ matrix, and
\begin{equation}
\hat{K}_{\vec{k}^\prime} = \left(\begin{array}{c c c}
k^{\prime 2}  & 0 & 4 i k_y ^\prime \\
0 & k^{\prime 2}  & -4 i k_x ^\prime \\
-4 i k_y ^\prime & 4 i k_x ^\prime  & k^{\prime 2} 
\end{array} \right) .
\end{equation}

Equation (\ref{Matrix_Eqn}) can be solved analytically for $\vec{M}_{\vec{k}}$ via matrix inversion,
\begin{equation}
\vec{M}_{\vec{k}} = \sum_{\vec{k}^\prime} \hat{D}_{\vec{k}\vec{k}^\prime} ^{-1} \vec{I}_{\vec{k}^\prime} , \label{MatrixEqSol}
\end{equation}
yielding a magnetization functional $\vec{M}_{\vec{k}}[\lambda_{\vec{k}}]$ that depends on the Lagrange multiplier field. This solution, however, does not necessarily obey the micromagnetic constraint $| \vec{M}(x,y) |^2 = 1$. In order to enforce it, we must then solve the constraint equation numerically to find the appropriate Lagrange multiplier. In Fourier space, this amounts to solving the following equations for $\lambda_{\vec{k}}$,
\begin{align}
\sum_{\vec{k}} | \vec{M}_{\vec{k}} |^2 & = 1, \\
\sum_{\vec{k}} \vec{M}_{\vec{k}} \cdot \vec{M}_{\vec{q} - \vec{k}} & = 0, 
\quad \forall \vec{q} \neq 0,
\end{align}
where we use the $\lambda$-dependence of the magnetization, $\vec{M}_{\vec{k}} = \vec{M}_{\vec{k}}[\lambda_{\vec{k}}]$, from Eq.~(\ref{MatrixEqSol}). 
We find that the Lagrange multiplier $\lambda(x,y)$ takes the largest values in regions where the spins are aligned parallel or anti-parallel to the external field and the lowest values in regions where the spins are perpendicular to it (parallel to the $xy$-plane). 

\subsection{\label{sec:SkX} Skyrmion Lattice}

Typical solutions obtained by the procedure described in the previous section are shown in Fig.~\ref{Stable}. 
Keeping in mind that the applied field is directed in positive $z$-direction, a typical skyrmion lattice solution appears with spins aligning along the field direction, $z$, in the region between skyrmions, and in the opposite direction in the centers of skyrmions. 
Hence, the region between skyrmions is mostly ferromagnetic, while in moving away from the center of a skyrmion along any radial direction, the spin orientation rotates continuously within the plane perpendicular to the given radial direction, acquiring a component along the $x$-$y$ plane, in a similar fashion as found in helical magnetic patterns. In this way, the system is able to gain energy from the DM interaction, which favors the twisting of spins, while retaining a large region of ferromagnetic alignment, promoted by the Zeeman energy.

\begin{figure}[t!]
\centering
\includegraphics[width=0.22\textwidth]{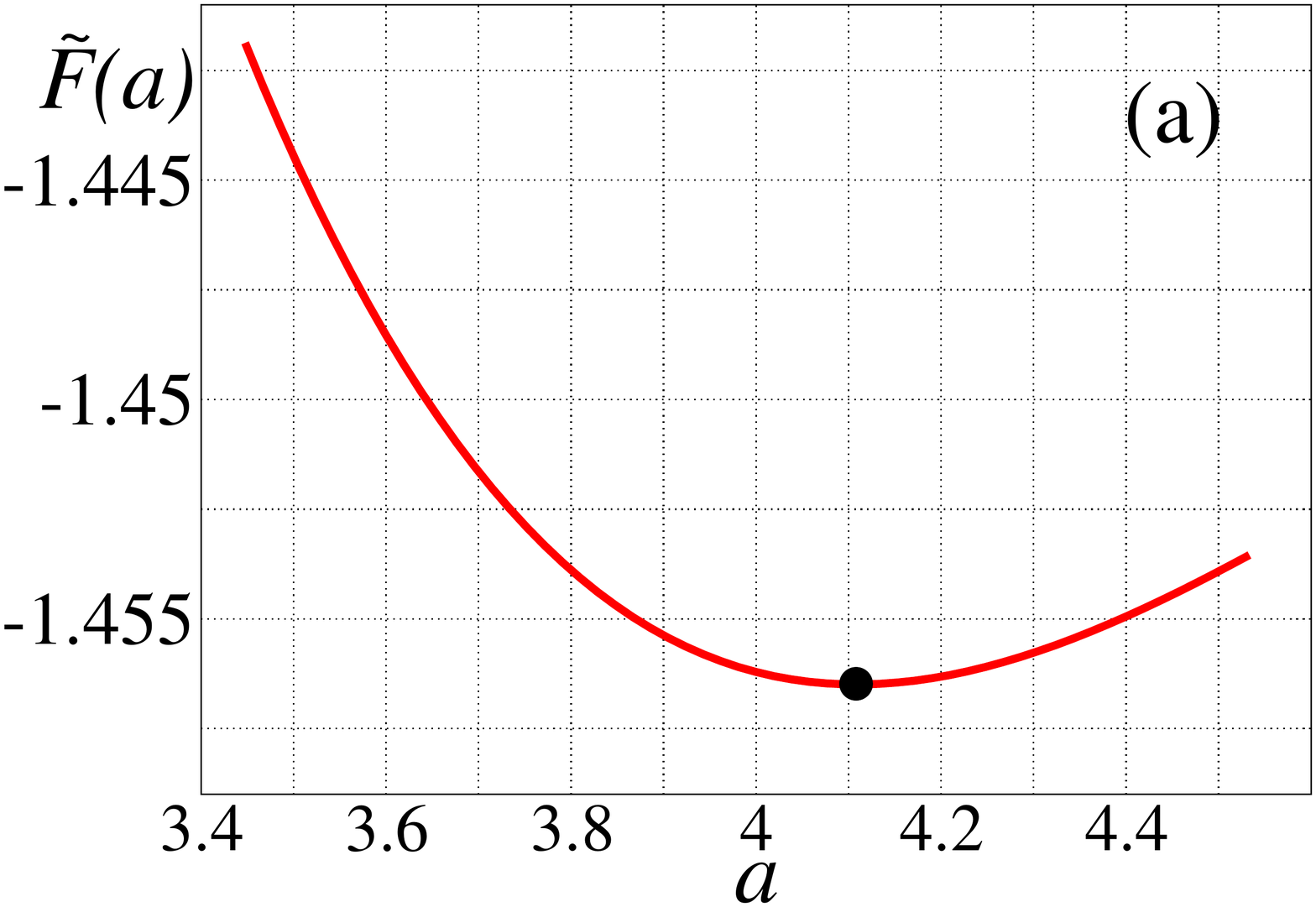}
\includegraphics[width=0.22\textwidth]{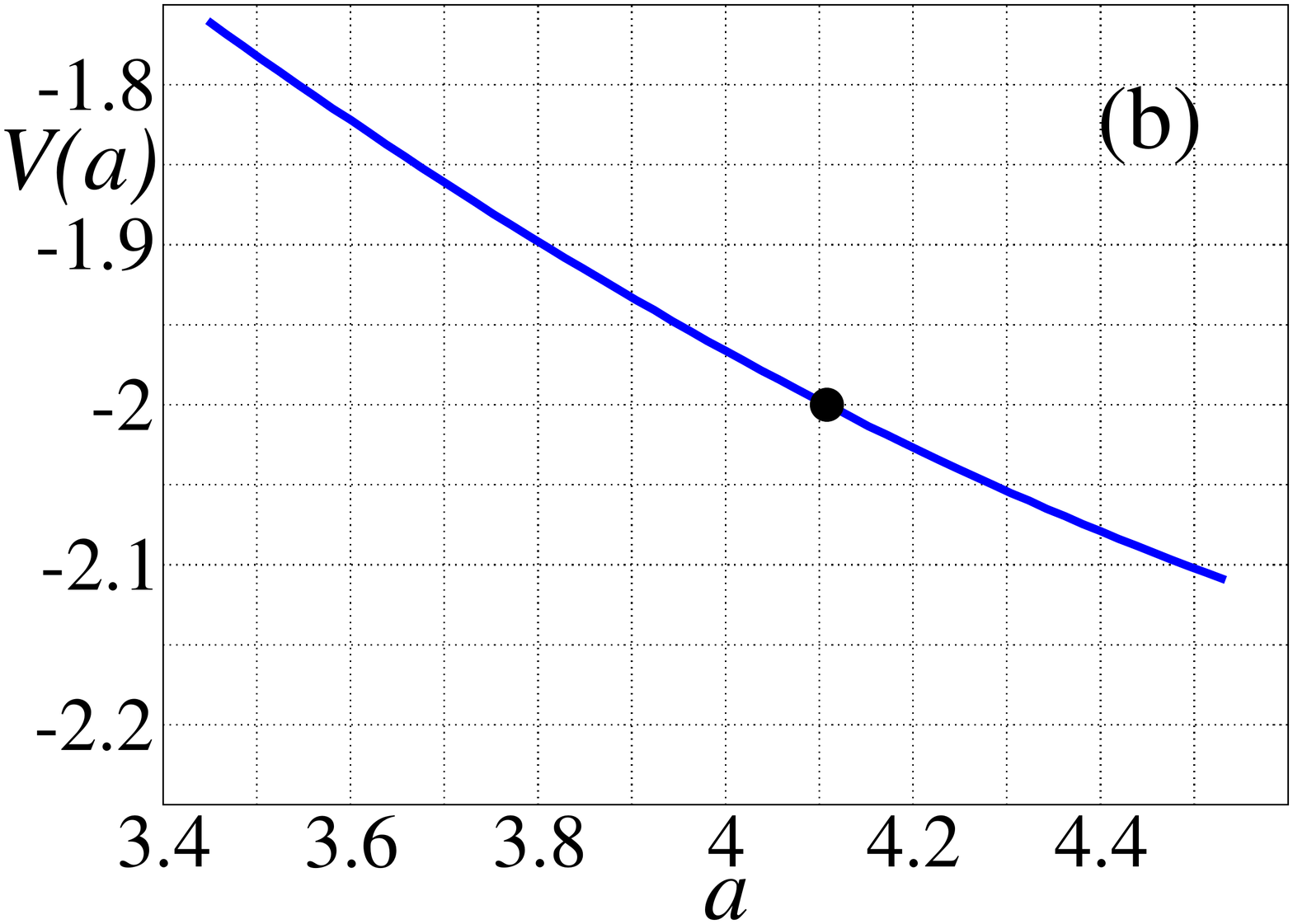}
\caption{(a) Free energy $\tilde F$ and (b) virial ratio $V=\tilde{F}_{DM}/\tilde{F}_{ex}$,
both as function of skyrmion lattice constant $a$, for $\beta = 1.4$. The black dot indicates the minimum of the free energy in (a) and the point corresponding to $V = -2$ in (b).}
\label{Energy_Spacing}
\label{Virial_Spacing}
\label{S_E_V}
\end{figure}

Shown in Fig.~\ref{Stable} are two self-consistently determined solutions for a triangular skyrmion lattice, one close to the phase transition to a spiral state (top panels), and one close to the phase transition to the homogeneous ferromagnetic state (bottom panels).
In the first case the spatial modulation of the magnetization vector along a path going from one skyrmion center to a neighboring one approaches that of a spiral modulation, whereas in the latter case the skyrmions become isolated from each other and the inter-skyrmion regions expand. It is also interesting to observe, in these two extreme cases, the different spatial distribution of the $z$-component of the magnetization within a single skyrmion, whose radius can be conventionally identified in the red dashed contour ($M_z = 0.95$). In the first case the area of negative magnetization (blue region, enclosed by the thick black contour) is approximately equal to the area of positive magnetization (red region, enclosed between the red dashed contour and the thick black contour). Hence, skyrmions in this case carry, as a whole, essentially no net magnetic moment, as in the case of a helix. On the other hand, in the vicinity of the ferromagnetic transition the area of positive magnetization is nearly three times that of negative magnetization, so that skyrmions do carry a finite magnetic moment along the external field direction. As we shall see later on, the shrinking of the negative magnetization region plays an important role in determining the magnetic field dependence of the inter-skyrmion distance.

Solutions to the Fourier transformed Euler-Lagrange equations, (\ref{EL_Fourier_1})--(\ref{EL_Fourier_3}), satisfy the condition of stationarity of the functional within a given lattice geometry. However, there is no guarantee that such a solution is truly stationary with respect to arbitrary variations of the magnetization pattern and, even if this is the case, that the solution represents a global minimum of the functional and not just a local one. Hence, it is necessary to devise an additional procedure to distinguish stable solutions from those metastable or not stable at all. 

We recall that the method described above is based on a Fourier expansion of the functional, where quantities such as the magnetization and the Lagrange multiplier are assumed to be periodic on a given lattice structure (triangular, for the time being) with lattice spacing $a$. When seeking a solution to the Euler-Lagrange equations we must therefore remember that we are only exploring the subspace of magnetization profiles with a fixed given periodicity. The classes of solutions that we can obtain in this way include (but are not limited to) the helical magnetic order, the triangular skyrmion lattice, and the ferromagnetic order. Hence, in order to find the globally stable solution, i.e., the ground state of the system, we must first minimize, with respect to the lattice spacing $a$, the free energy corresponding to a given class of solutions, and later on compare the optimal energies of each class of solutions, identifying the lowest one.

A typical profile of the free energy as a function of the lattice spacing is shown, for the skyrmion lattice class of solutions, in Fig.~\ref{Energy_Spacing} (a). The free energy potential $\tilde{F}(a)$ has a distinct minimum, corresponding to the optimal spacing of a triangular skyrmion lattice. The solutions shown in Fig.~\ref{Stable} are calculated for such free energy minima.

Although the value of the optimal spacing can be, in principle, extracted from this curve alone, one must be aware that its precise determination requires the evaluation of the potential at many points in $a$ around the minimum, due to the quadratic dependence of $\tilde{F}(a)$. On the other hand, there is a far more efficient procedure to extract the optimal spacing, and it is based on the virial theorem. Even more importantly, the virial theorem method allows one to discern whether or not the optimal spacing solution for a given lattice is truly stationary with respect to variations in the lattice geometry.

\subsection{\label{sec:virial} Virial theorem}

In order to formulate the virial theorem for our problem, we rescale the coordinates inside the spatial integrals via a dimensionless parameter $\gamma$, writing $\vec{r} = \gamma \vec{r}^{\,\prime}$. This leads to the following rescaling of the dimensionless free energy, Eq.~(\ref{Functional_dimless}),
\begin{align}
\tilde{F}\left[\vec{M} \right] & = \tilde{F}_{ex}\left[\vec{M} \right] 
 + \tilde{F}_{DM}\left[\vec{M} \right] 
 + \tilde{F}_B\left[\vec{M} \right] \label{eq:rescaling} \\
 & = \frac{1}{\gamma^2} \tilde{F}_{ex}\left[\vec{M}_\gamma \right] 
 + \frac{1}{\gamma} \tilde{F}_{DM}\left[\vec{M}_\gamma \right] 
 + \tilde{F}_B\left[\vec{M}_\gamma \right] , \nonumber
\end{align}
where $\tilde{F}_{ex}$ is the ferromagnetic exchange term, $\tilde{F}_{DM}$ is the DM term, $\tilde{F}_B$ is the Zeeman term, and 
$\vec{M}_\gamma(\vec{r}^{\,\prime}) = \vec{M}(\gamma \vec{r}^{\,\prime})$. 
Since the above rescaling is just the result of a change of variables in the integrals, Eq.~(\ref{eq:rescaling}) holds for any value of $\gamma$. Hence, the right-hand side of Eq.~(\ref{eq:rescaling}) must be independent of $\gamma$ and, accordingly, its derivative with respect to $\gamma$ must vanish, 
\begin{align}
 0 & = -2 \frac{1}{\gamma^3}\tilde{F}_{ex}\left[\vec{M}_\gamma \right] 
 - \frac{1}{\gamma^2}\tilde{F}_{DM}\left[\vec{M}_\gamma \right] \label{gamma_stuff} \\
 & + \int \left[ 
 \frac{1}{\gamma^2} \frac{\delta \tilde{\mathcal{F}}_{ex}}{\delta \vec{M}_\gamma} 
 + \frac{1}{\gamma} \frac{\delta \tilde{\mathcal{F}}_{DM}}{\delta \vec{M}_\gamma}
 + \frac{\delta \tilde{\mathcal{F}}_{B}}{\delta \vec{M}_\gamma}
 \right] \frac{\partial \vec{M}_\gamma}{\partial \gamma} 
 \frac{dx^\prime dy^\prime}{A^\prime} . \nonumber
\end{align}
We can now use the arbitrariness in the value of $\gamma$ and set $\gamma=1$ in the above equation, in order to obtain the following relation between free energy terms,
\begin{equation}
 -2 \tilde{F}_{ex}\left[\vec{M} \right] 
 - \tilde{F}_{DM}\left[\vec{M} \right] 
 + \int 
 \frac{\delta \tilde{\mathcal{F}}}{\delta \vec{M}} (\vec{r} \cdot \nabla) \vec{M}(\vec{r})
 \frac{dx dy}{A} = 0 , \label{virial_generic}
\end{equation}
where we have used 
$\left. \frac{\partial \vec{M}_\gamma}{\partial \gamma} \right|_{\gamma=1} 
= (\vec{r} \cdot \nabla) \vec{M}(\vec{r})$.
The above equation holds for any magnetization profile $\vec{M}(\vec{r})$, which need not be a minimum of the free energy functional. However, 
when $\vec{M}(\vec{r})$ represents a stationary solution of the Euler-Lagrange equations, the additional condition
\begin{equation}
\frac{\delta \tilde{\mathcal{F}}}{\delta \vec{M}} = 0 
\end{equation}
holds.
Therefore, when evaluated on a genuinely stationary solution $\vec{M}_{\rm stat}(\vec{r})$, Eq.~(\ref{virial_generic}) reads 
\begin{equation}
-2 \tilde{F}_{ex}[\vec{M}_{\rm stat}] - \tilde{F}_{DM}[\vec{M}_{\rm stat}] = 0 .
\end{equation}
This is the virial theorem \cite{Doria89} formulated for our problem, \cite{Bogdanov94b,Leonov16} which can be rewritten in terms of the following ``virial ratio'',
\begin{equation}
\frac{\tilde{F}_{DM}}{\tilde{F}_{ex}} = - 2 , \label{Additional_condition_0}
\end{equation}
satisfied by stationary magnetization profiles.

In Fig.~\ref{Virial_Spacing} (b) we plot the virial ratio (\ref{Additional_condition_0}) against the lattice spacing. As expected, the minimum of the free energy potential coincides precisely with the value of spacing for which we have ${\tilde{F}_{DM}}/{\tilde{F}_{ex}} = -2$, indicating that the solution is a genuinely stable configuration. Furthermore, we clearly see that the virial ratio has a linear dependence on the lattice spacing, and this allows for a much more efficient and precise numerical determination of its optimal value $a_{\rm opt}$, which we can now define through the condition $({\tilde{F}_{DM}}/{\tilde{F}_{ex}})|_{a_{\rm opt}} = -2$, rather than through the minimum of the potential.

\subsection{\label{sec:level3} Phase Diagram and Critical Points}

The free energy can be expressed in terms of the Fourier components via
\begin{eqnarray}
\tilde{F} = & & \sum_{\vec{k}} \left\lbrace \frac{1}{4} k^2 \left(\left|X_{\vec{k}} \right|^2 + \left|Y_{\vec{k}} \right|^2 + \left|Z_{\vec{k}} \right|^2 \right) \right. \\
&+& \left. i \left( k_y X_{\vec{k}} Z_{\vec{k}} ^* - k_x Y_{\vec{k}} Z_{\vec{k}} ^* + k_x Y_{\vec{k}} ^* Z_{\vec{k}} - k_y X_{\vec{k}} ^* Z_{\vec{k}} \right)\right\rbrace \nonumber \\ 
&-& \beta Z_0 . \nonumber \label{Functional_Fourier_Full}
\end{eqnarray}
This expression allows for a fast and precise evaluation of the free energy without the need to evaluate numerical derivatives.

\begin{figure}[tb]
\centering
\includegraphics[width=0.5\textwidth]{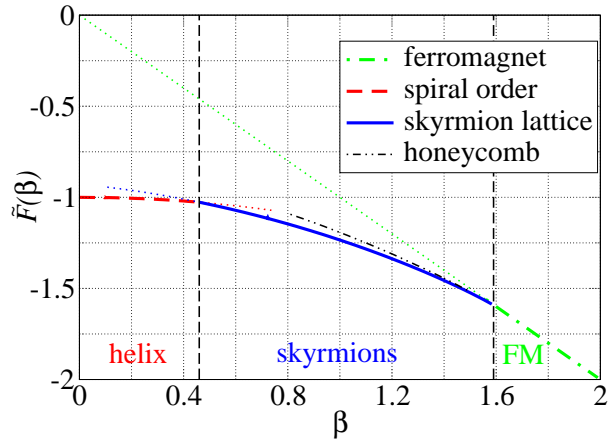}
\caption{
Free energy $\tilde F$ of a magnetic system as function of external field $\beta $. 
The full curve represents the optimal free energy of a skyrmion lattice (which is triangular in its ground state), the dashed line represents a system with spiral order, and the dot-dashed line a system with homogeneous ferromagnetism. Dotted curves represent metastable extensions into neighboring phases, resulting from local, but not global, minima in the free energy. A honeycomb skyrmion lattice solution, indicated by the black dashed-double-dot line, is possible within the subspace of lattices with trigonal symmetry.}
\label{Metastable1}
\end{figure}

In the following, we discuss solutions minimizing the free energy for different external magnetic fields, starting from $\beta = 0$. In order to get a better understanding of the transitions, it is also instructive to include the free energy functional of metastable solutions next to the stable ones. This is shown in Fig.~\ref{Metastable1}, where energies of metastable helices, skyrmions, and ferromagnetic state are plotted alongside the stable solutions.

Figure~\ref{Metastable1} also shows the critical values of $\beta $ for phase transitions. 
In our units the critical field for the helical-skyrmion phase transition is $\beta_{c_1} = 0.46$ and the critical field for the skyrmion-ferromagnet phase transition is $\beta_{c_2} = 1.59$. This is in excellent agreement with the experimentally observed phase diagram of the 2D $\rm{Fe_{1-x}Co_xSi}$ chiral magnet,\cite{Yu10} and with previous theoretical investigations such as Monte-Carlo simulations \cite{Yi09} and analytical approximate methods.\cite{Han10} 
We find that the helical-skyrmion transition at $\beta_{c_1}$ is clearly first-order, as underlined by the possibility to extend, as metastable configurations, the skyrmion or spiral order into the neighboring phase (blue and red dotted lines in Fig.~\ref{Metastable1}). The skyrmion-ferromagnet transition at $\beta_{c_2}$, instead, appears to be second-order in the continuum limit. This is suggested by the impossibility to find metastable skyrmion lattices above $\beta_{c_2}$ and by the diverging behavior of the inter-skyrmion distance in approaching the transition, as discussed hereafter. However, we have to keep in mind that the continuum limit is only an approximation and that in any actual material the spins originate from a crystal lattice. Hence, for real materials the second-order transition could easily turn into a weakly first-order one.

\begin{figure}[!t]
\centering
\includegraphics[width=0.5\textwidth]{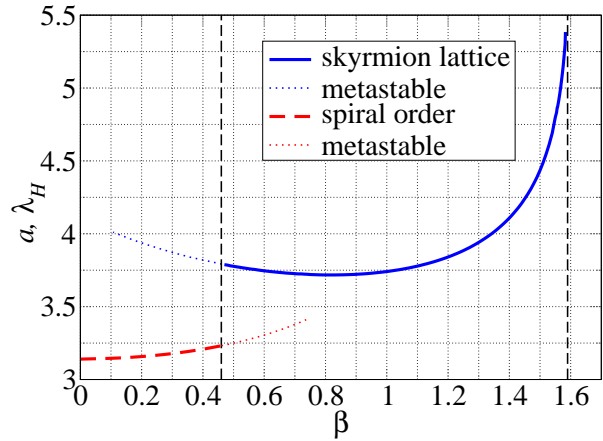}
\caption{
The full curve is the lattice constant $a$ of the skyrmion lattice as function of $\beta$, the dashed curve is the helical length $\lambda_H$ as function of $\beta $ for the spiral order, and the dotted curves are metastable extensions.
The critical fields $\beta_{c_1}$ and $\beta_{c_2}$ are shown as vertical dashed lines.
The spacing increases drastically when the field approaches the skyrmion-ferromagnet transition, $\beta_{c_2}$. The skyrmion lattice spacing also slightly increases when the field approaches the skyrmion-helical transition, $\beta_{c_1}$, from above. }
\label{Spacing}
\end{figure}
	\begin{figure*}[!tb]
	\centering
	{\includegraphics[width=0.45\textwidth]{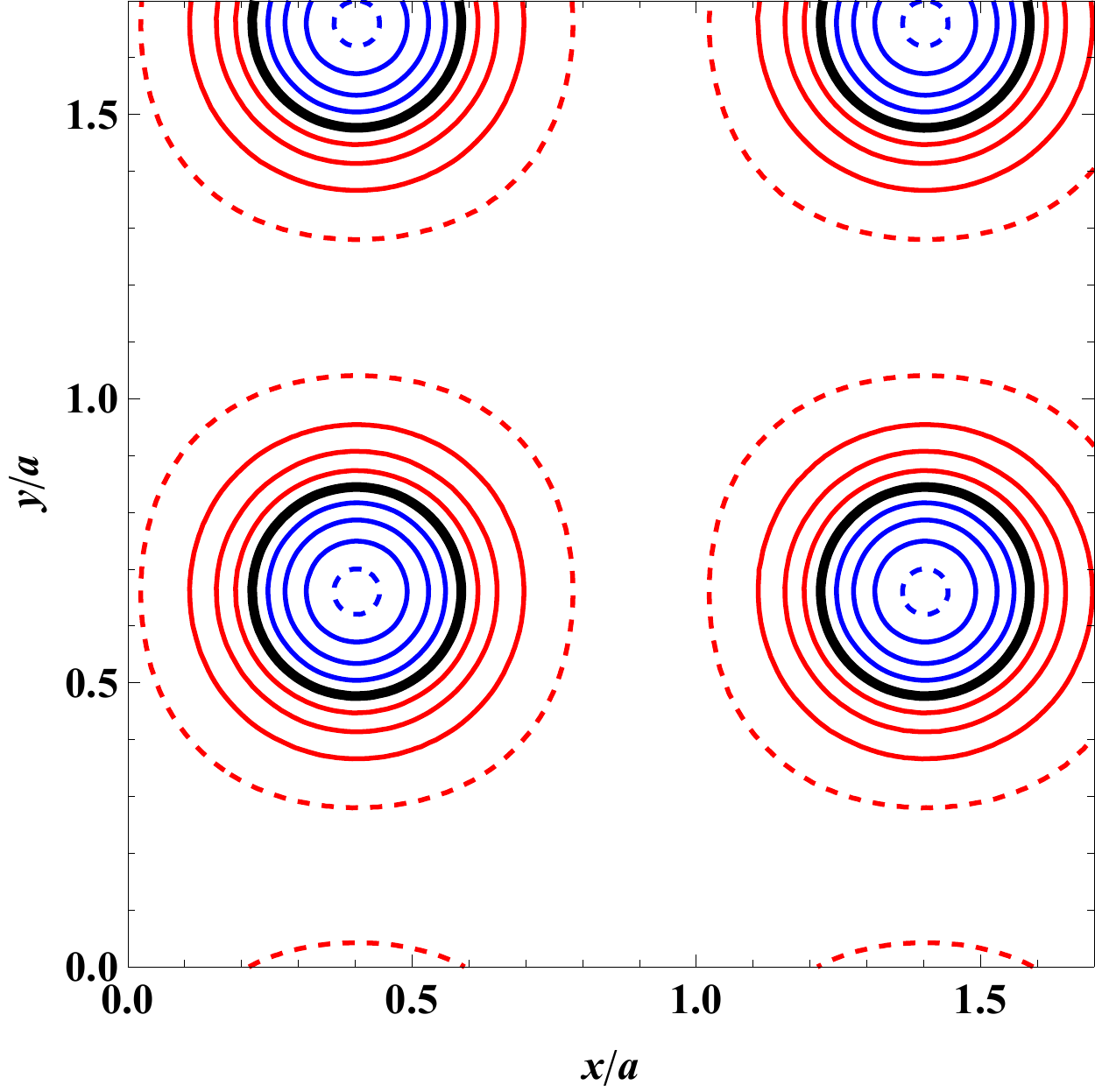}} \hfill
	{\includegraphics[width=0.45\textwidth]{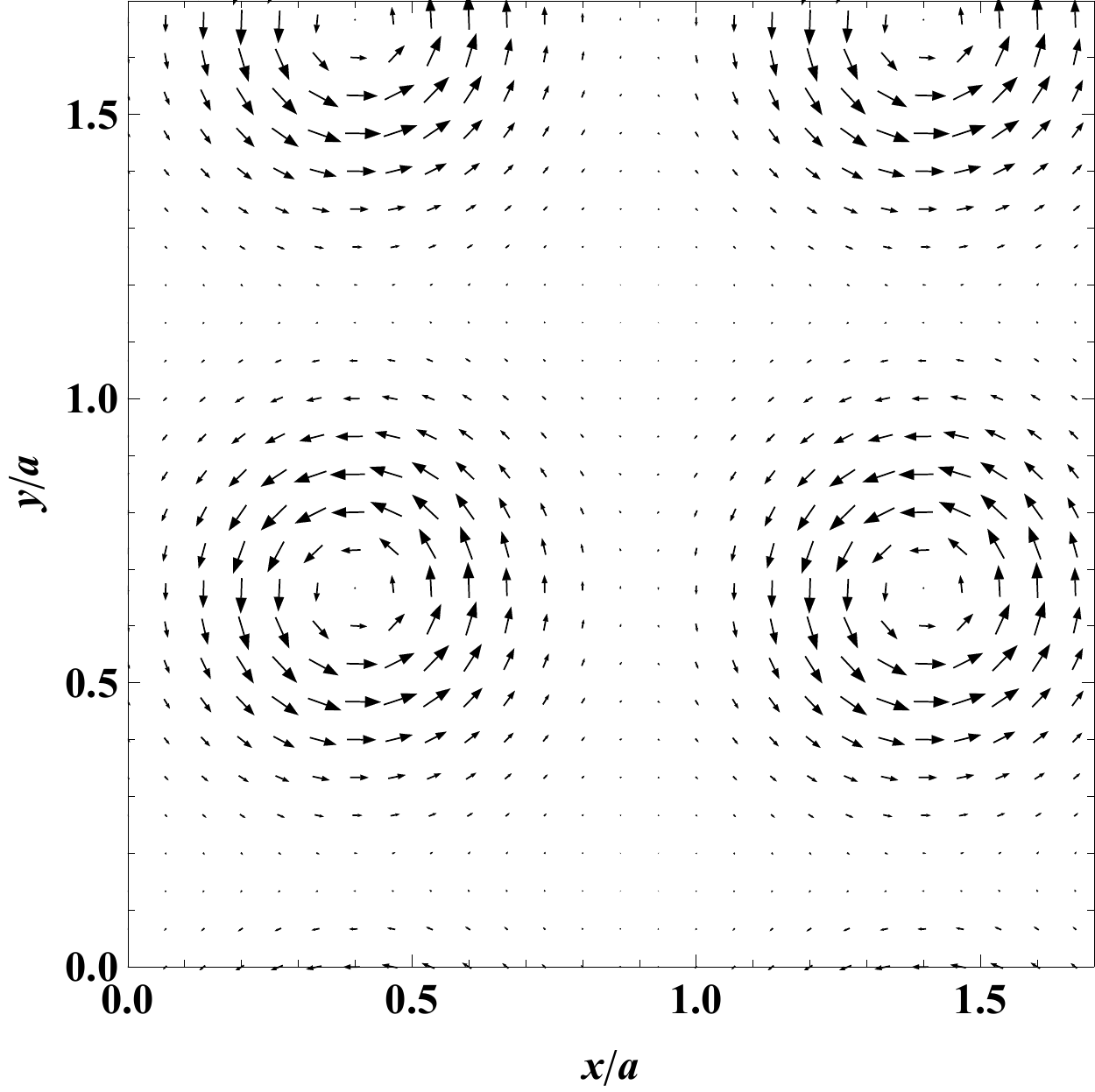}}
\\[0.5cm]
	{\includegraphics[width=0.45\textwidth]{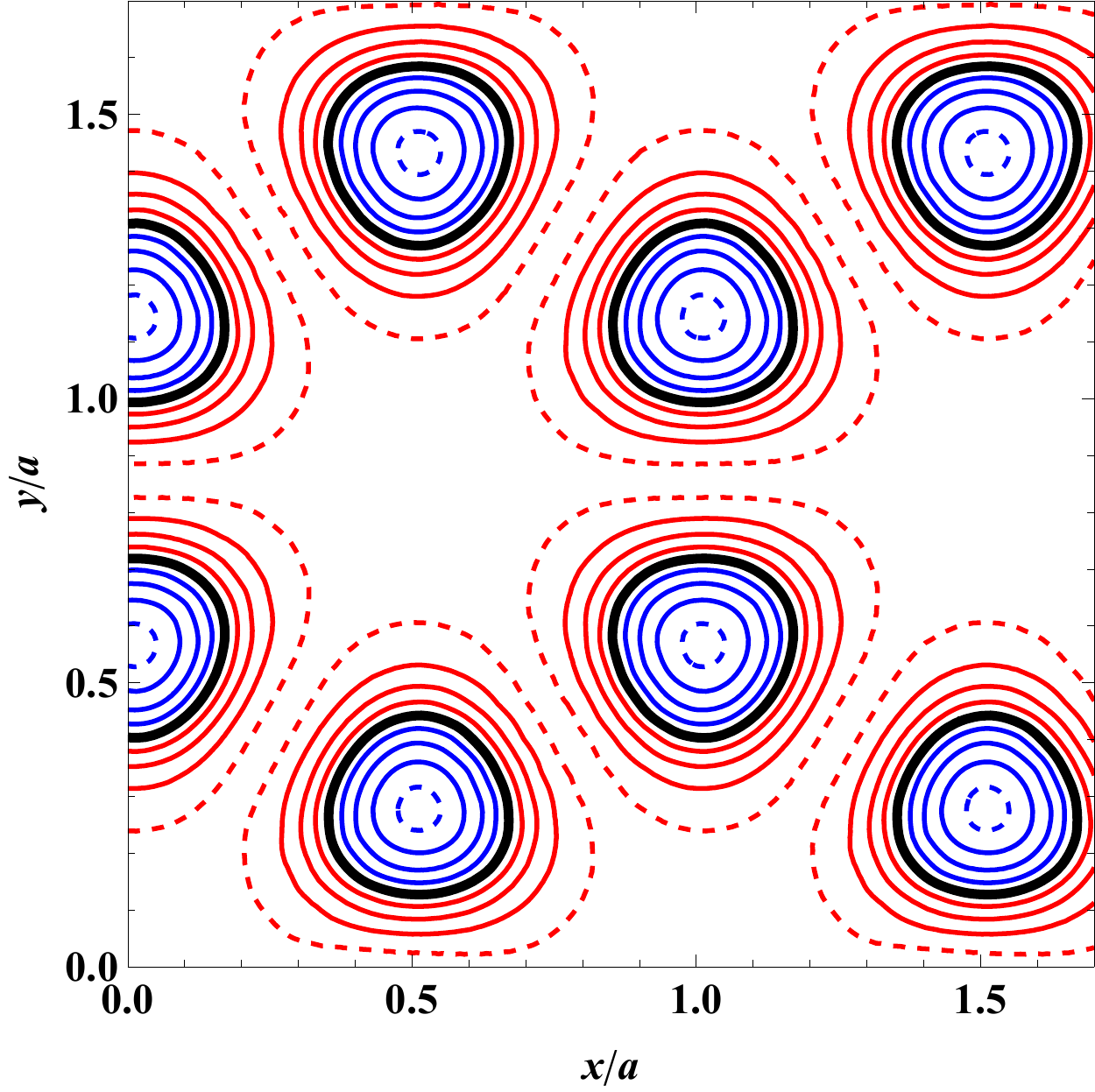}} \hfill
	{\includegraphics[width=0.45\textwidth]{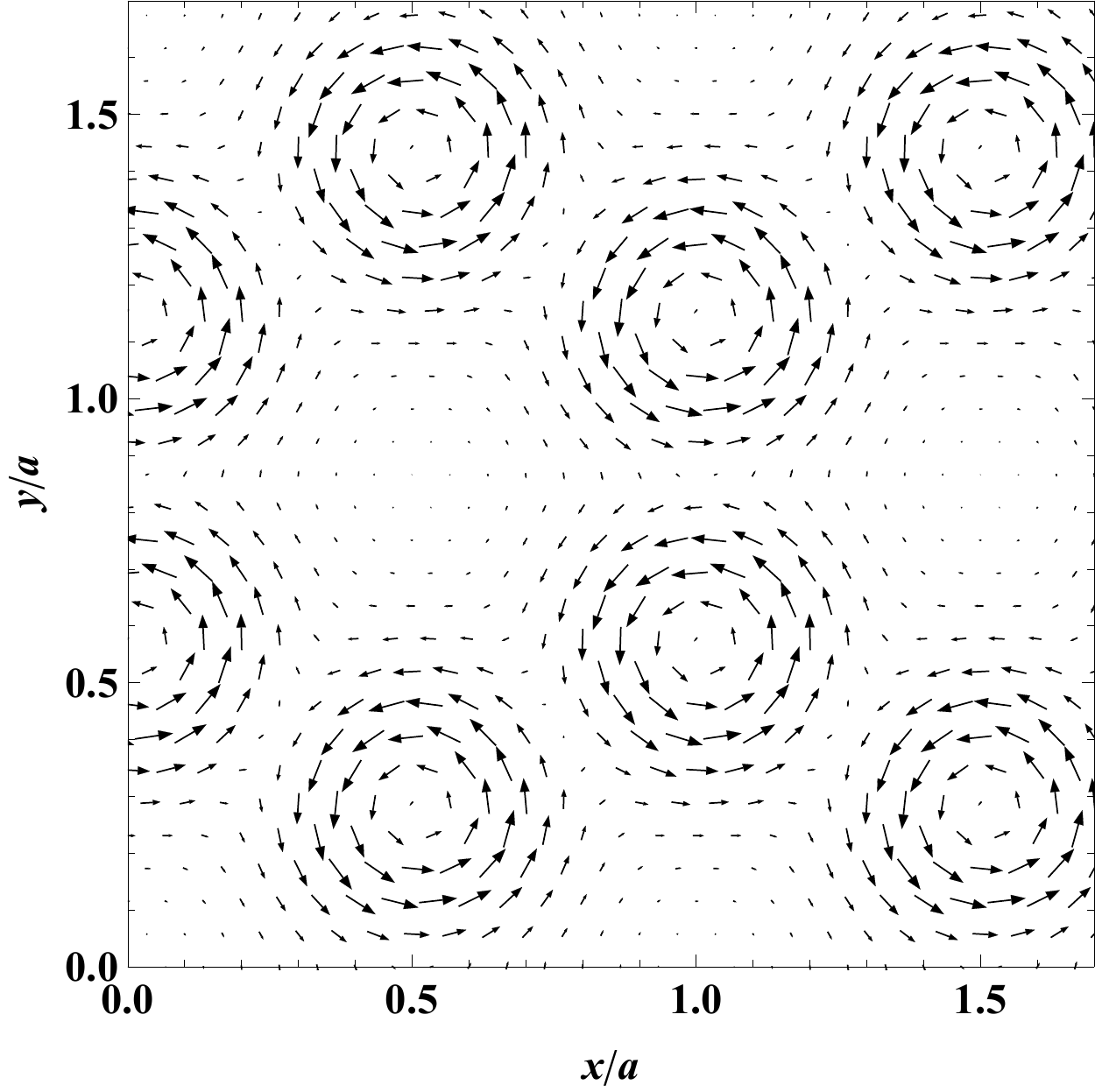}}
	\caption{
Top panels: Metastable square skyrmion lattice solution for an external field of $\beta = 1.5$ and a lattice spacing of $a = 4.2$.
Bottom panels: Metastable honeycomb skyrmion lattice solution for an external field of $\beta = 1.031$ and a lattice spacing of $a = 5.734$; this solution is metastable only within the subset of lattices with trigonal symmetry. 
Left: contour plots for $M_z$, with the contour lines at values $-0.95$ (blue dashed), $-0.75, -0.5, -0.25,$ 0 (thick line), 0.25, 0.5, 0.75, 0.95 (red dashed), with negative values in blue and positive values in red. The centers of the skyrmions have $M_z=-1$. Right: the corresponding vector fields $(M_x,M_y)$, showing the anti-clockwise curling of the magnetization vector for each skyrmion in the lattice.
All spatial coordinates are normalized to the (self-consistently determined) lattice spacing $a$.}
	\label{Square}
	\label{Honeycomb}
	\end{figure*}

The lattice spacing that provides an energetically optimal configuration varies with the external magnetic field, a behavior that recalls the analogy with Abrikosov vortex lattices in type-II superconductors. However, while the lattice spacing of a superconducting vortex lattice is a monotonic function of the external field, scaling as the inverse square root of the magnetic induction and being determined by the magnetic flux quantization,\cite{Abrikosov57} 
the optimal spacing of a skyrmion lattice, shown in Fig.~\ref{Spacing},
exhibits a non-monotonic dependence upon the external field.
Starting from the helical-skyrmion transition, the top panels of Fig.~\ref{Stable} indicate that in the vicinity of $\beta_{c_1}$ skyrmions are close packed, so that the lattice spacing is roughly twice the skyrmion radius. Moreover, as already discussed in Section~\ref{sec:SkX}, in this region a given skyrmion is characterized by approximately an equal amount of negative and positive $M_z$, resulting in a zero net magnetic moment. If we now increase the magnetic field, the area of negative magnetization around the core of the skyrmion starts to shrink with respect to the positive one, giving the skyrmions a finite net magnetic moment along the field direction, and at the same time reducing the skyrmion radius. Since skyrmions at this stage are still close packed, the slightly smaller skyrmion radius results in a smaller lattice spacing, which reaches a minimum at $\beta = 0.82$. Upon further increasing the magnetic field, the skyrmion radius slowly continues to decrease. However, skyrmions cease to be close packed and become more and more isolated, separated from each other by increasingly large ferromagnetic domains, as shown in the bottom panels of Fig.~\ref{Stable}. Accordingly, the skyrmion lattice spacing increases rapidly when moving towards the ferromagnetic phase. Exactly at the $\beta_{c_2}$ transition, the behavior shown in Fig.~\ref{Spacing} seems to indicate a divergent lattice spacing, signaling a continuous vanishing of the skyrmion density. This scenario would imply a second-order phase transition, as mentioned earlier. Unfortunately, with our calculations we are not able to assert this with complete certainty due to the finite truncation of the Fourier series required by the numerical implementation. In fact, Eq.~(\ref{eq:k-vectors}) shows that, for an increasingly large lattice spacing $a$, the discrete $\vec{k}_{mn}$ vectors defining the Fourier series become closer and closer to each other, so that to achieve the same spatial resolution one has to include more and more $\vec{k}_{mn}$ vectors. In our numerical implementation, instead, we limited the number of $\vec{k}_{mn}$ vectors to a $25 \times 25$ grid.

In addition to the skyrmion lattice spacing, in Figure~\ref{Spacing} we present the field dependence of $\lambda_H = \frac{2\pi}{k_H}$, the period of a helix. At zero magnetic field one can prove analytically that $k_H = 2$ (in dimensionful units, $k_H = 2 \kappa = D/J$), so that the optimal value for a helix is \cite{Dzyaloshinsky58, Dzyaloshinskii64} $\lambda_H = \pi$. From the low-$\beta$ branch of the curve in Fig.~\ref{Spacing} we can see that, at $\beta = 0$, $\lambda_H$ obtained from our calculation is exactly equal to $\pi$, showing that our calculations match the analytical predictions. As long as the field is turned on, the helix gets deformed, rotating faster in the regions where the magnetization is opposite to the field direction, and slower where the magnetization points along the field, so that the system can gain some Zeeman energy from the latter regions. Overall, this leads to a larger period of the helix, which thereby increases with the external field.

\section{\label{sec:Meta} METASTABLE SKYRMION LATTICES}

In addition to the stable triangular skyrmion lattice phase illustrated above, we shall now discuss a few examples of alternative metastable skyrmion lattice geometries, and a possible way to stabilize them. If for a stable triangular lattice the virial ratio, and specifically its linear dependence on the lattice spacing, provided a numerically efficient and accurate tool to find the optimal spacing, in the case of metastable solutions it plays an even more important role. In fact, for lattice geometries different from the triangular one, there is no guarantee that the solution obtained by minimizing the free energy with respect to the lattice spacing represents a true metastable solution. Changing the lattice spacing is indeed only one of the possible ways to vary the underlying lattice structure, the others being a change in the ratio and in the angle between the unit vectors of the lattice---namely, a change in the lattice geometry. As discussed in Section~\ref{sec:virial}, any truly stationary solution of the Euler-Lagrange equations, including metastable skyrmion lattices, should fulfill the virial theorem. Hence, we can use this criterion to distinguish true metastable solutions from those that are local minima of the free energy only within the subset of solutions with a given lattice symmetry. 

The first type of metastable skyrmion lattice that we are going to discuss is a square lattice, shown in the top panels of Fig.~\ref{Honeycomb}. This pattern can be realized, as an alternative to the triangular lattice, in the whole skyrmions phase between $\beta_{c_1}$ and $\beta_{c_2}$, and it represents a genuine metastable solution in the sense that it is a local minimum of the free energy functional with respect to arbitrary variations of the magnetization pattern. This is confirmed by the virial ratio, which takes the value of $-2$ at the minimum of the free energy as a function of the lattice spacing. 
In the phase diagram of Fig.~\ref{Metastable1}, the energy of the square lattice (not shown to avoid an overburdening of the plot) is only slightly higher than the ground state energy, 
lying between the triangular (blue full curve) and honeycomb (black double-dot-dashed curve) skyrmion lattice energies. In approaching the skyrmion-ferromagnet transition, all skyrmion energies merge together at $\beta_{c_2}$. Such a behavior is precisely what one would expect from a second-order phase transition, characterized by a vanishing skyrmion density (or, equivalently, a diverging lattice spacing). Indeed, it is clear that for nearly isolated skyrmions the details of the lattice symmetry become unimportant and the energy of the system is controlled only by the skyrmion density.

\begin{figure}[t!]
\centering
\includegraphics[width=0.48\textwidth]{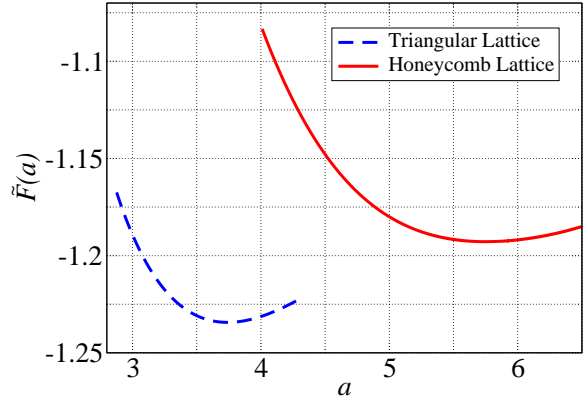}
\caption{
Free energy of a honeycomb configuration compared with
the free energy of a triangular configuration as function of 
lattice spacing $a$ for a fixed value of $\beta=1.0$.}
\label{Energy_spacing_honeycomb}
\end{figure}

The second type of metastable phase that we would like to present is, instead, a peculiar metastable solution that was found while investigating the system, namely a honeycomb skyrmion lattice, depicted in the bottom panels of Fig.~\ref{Honeycomb}. As it is well known for honeycomb lattices, the underlaying Bravais lattice for this solution is still a triangular lattice, however the unit cell contains now two skyrmions and the lattice spacing is no longer the distance between the centers of nearest skyrmions, but the distance between the centers of honeycombs. In Fig.~\ref{Energy_spacing_honeycomb} we show the free energy of a honeycomb skyrmion lattice as a function of the lattice spacing for a given $\beta$. As in the case of a triangular lattice (shown in the same figure for comparison), the free energy has a clear minimum at a certain optimal spacing. However, in stark contrast with the triangular and square lattice solutions presented earlier, the honeycomb lattice does not fulfill the virial theorem at the minimum of the corresponding free energy. This behavior entails a strong message, telling us that this solution cannot be a true local minimum of the free energy functional. Indeed, the free energy of such a configuration does have a local minimum as a function of the lattice spacing, but only as long as one considers lattice configurations with trigonal symmetry. Instead, if the geometry of the lattice is allowed to change, this solution can relax into a lower energy configuration. At first, one might think that configurations with high symmetries, such as the square or trigonal one, should be protected against deformations of the lattice geometry, so that the corresponding magnetization patterns should always be at least stationary points of the free energy functional. However, this is true only for simple Bravais lattices such as the square or triangular lattices, but no longer holds in the case of Bravais lattices with a basis such as the honeycomb one. In the latter case, for example, one realizes that the honeycomb lattice can be smoothly turned into a square lattice, which has a lower energy, by modifying the angle and ratio between the unit vectors in a certain direction. A modification of the angle and ratio in the opposite direction, instead, leads to a very different configuration with a higher energy. The honeycomb lattice would then relax into the square lattice if the lattice symmetry is left unconstrained. On the other hand, it still represents a valid metastable solution if the space of solutions is constrained, by some external mechanism, to the subset of configurations with trigonal symmetry.

\begin{figure}[t!]
\centering
\includegraphics[clip,width=0.5\textwidth]{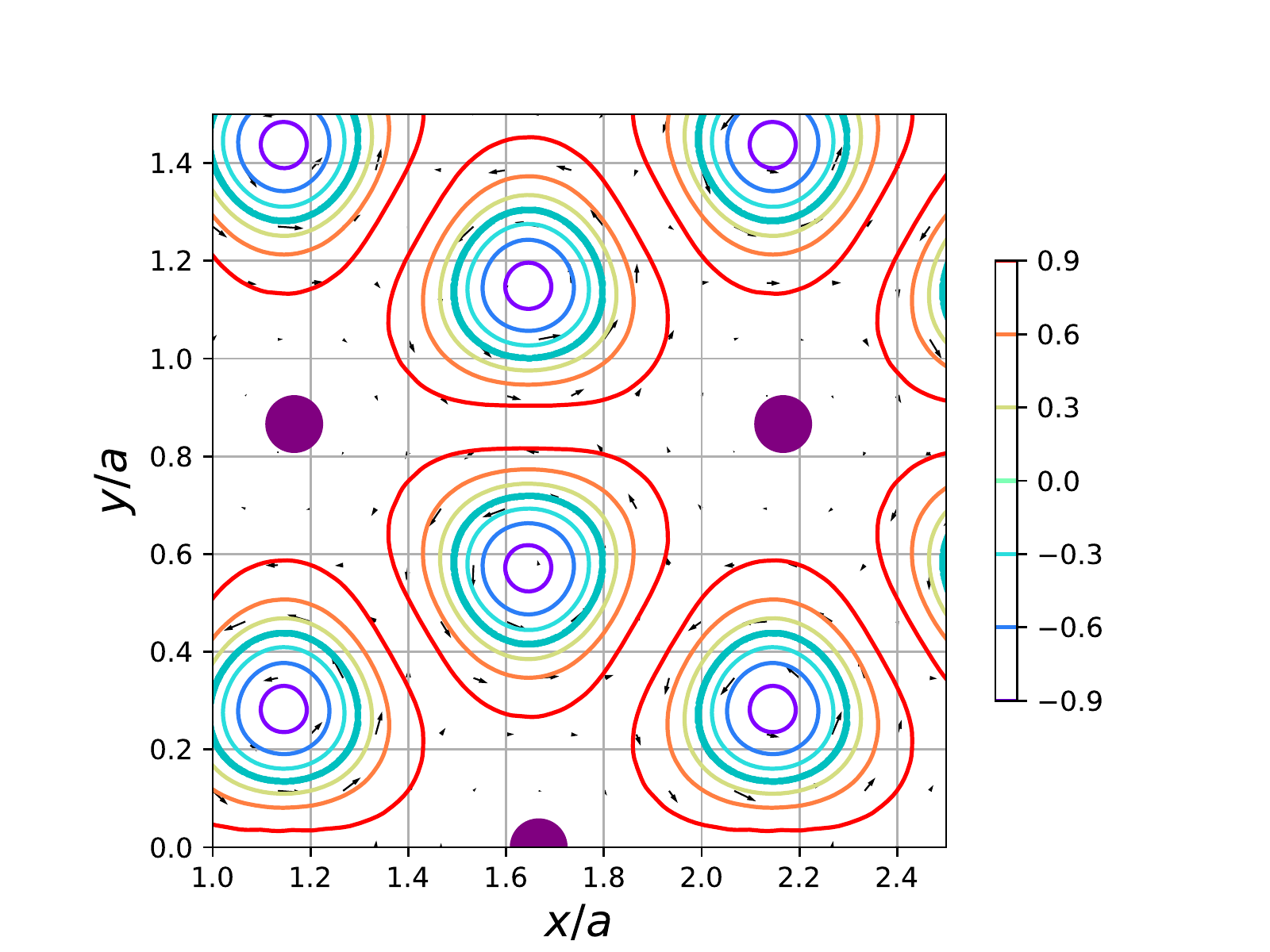}
\caption{Schematic representation of a honeycomb skyrmion lattice with Abrikosov vortices pinned at the centers of the honeycombs. }
\label{Honeycomb_vortices}
\end{figure}

As in the case of a square skyrmion lattice, the free energy of this peculiar metastable solution lies just above the free energy of the stable ground state configuration and, in approaching $\beta_{c_2}$, the difference between them becomes almost negligible. Hence, while this type of solution has not been experimentally observed in pure skyrmion materials, there is hope that the honeycomb skyrmion lattice might be stabilized by some sort of magnetic perturbation that establishes an underlying periodic structure with trigonal symmetry. In particular, by looking at the spin texture of this solution, we realize that skyrmions are arranged so as to surround and enclose large ferromagnetic domains. These domains are themselves arranged in a triangular lattice with vertices located at the centers of honeycombs. Since we already know that ferromagnetic domains are stabilized by a large magnetic field, a promising way of stabilizing this honeycomb skyrmion lattice is to make the spins experience an ``external'' magnetic field that is no longer constant, but modulated periodically so that it is largest at the vertices of a triangular lattice.

A noteworthy example of such a spatial pattern is given by the microscopic magnetic field of a type-II superconductor in the Abrikosov vortex phase, \cite{Abrikosov57} where the magnetic field is largest at the vortex cores and weaker in the interstitial superconducting region between vortices. Hence, provided that the length and magnetic field scales of the skyrmion material and type-II superconductor are compatible, we argue that a promising way to stabilize the honeycomb skyrmion lattice is to realize a bi- or tri-layer consisting of a non-centrosymmetric ferromagnet and a strong type-II superconductor, electromagnetically coupled, where the Abrikosov vortices would be pinned at the centers of the skyrmion honeycombs. Schematic image of such a system can be found in Fig.~\ref{Honeycomb_vortices}. A similar heterostructure has been very recently proposed, albeit with a different purpose, in Ref.~\onlinecite{Dahir19}, of which we became aware only after submitting our manuscript. There, the authors suggest that the coupling of a chiral ferromagnet with a type-II superconductor, characterized by an intertwined lattice of superconducting vortices and anti-vortices, might stabilize an ordinary triangular skyrmion lattice even in the absence of an external magnetic field, i.e., in the region where helical order is otherwise thermodynamically stable.

\section{\label{sec:Conc} SUMMARY}

In this paper we have introduced a numerically fast and stable approach for modeling magnetic systems by combining a Fourier-transform method with a virial theorem. We have demonstrated the accuracy of the method by applying it to a magnetic skyrmion lattice, reproducing the optimal skyrmion spacing and critical fields. 
Although we have considered only the case of a DM interaction induced by a Dresselhaus spin-orbit coupling, a Rashba type DM interaction, as well as easy-axis or easy-plane magnetic anisotropies, can be straightforwardly implemented in our method without incurring any additional numerical cost. Hence, we believe that our method should be perfectly suitable to describe also the recently predicted\cite{Lin15,Rowland16} and observed\cite{Yu18} skyrmion fractionalization into meron-antimeron lattices.

We then have discussed a novel metastable honeycomb lattice solution, and suggested to combine a skyrmion lattice with a superconducting vortex lattice in order to stabilize this solution. The new phase arranges superconducting  vortices in the inter-skyrmion regions, whereas skyrmions are placed in the regions where the local magnetic field is suppressed due to superconductivity. Both of these requirements can be fulfilled in a state hosting a skyrmion honeycomb lattice in combination with a triangular Abrikosov vortex lattice.
This system would also lead to interesting potential ways of manipulating skyrmions, which could be moved around by drifting the Abrikosov vortices.\cite{Rabinovich18,Reichhardt18,Becerra17} Moreover, the inhomogeneous magnetic pattern of skyrmions could be able to induce odd-frequency spin-triplet correlations in the superconductor already in a bilayer system, without the need of two separate ferromagnetic layers as in the case of superconducting spin valves.\cite{Kalenkov11,Yokoyama15}

Finally, the proximity between skyrmion textures and either s-wave or p-wave superconductors is a promising tool to generate and manipulate topologically non-trivial excitations such as Majorana bound states and tuneable Weyl points.\cite{Hals16,Yang16,Zhang16,Pershoguba16,Poyhonen16,Takashima16}

\acknowledgments
ME acknowledges support by EPSRC grant EP/N017242/1.

\end{document}